\documentclass[journal]{IEEEtran}
\ifCLASSINFOpdf

\else
 
\fi

\usepackage{cite}
\usepackage{amsmath,amssymb,amsfonts}
\usepackage{algorithmic}
\usepackage{graphicx}
\usepackage{textcomp}
\usepackage{longtable}
\usepackage{booktabs}
\usepackage{multirow}
\usepackage{subcaption}
\usepackage{dblfloatfix}
\usepackage{balance}

\def\BibTeX{{\rm B\kern-.05em{\sc i\kern-.025em b}\kern-.08em
    T\kern-.1667em\lower.7ex\hbox{E}\kern-.125emX}}

\begin{document}

\title{MAMAF-Net: Motion-Aware and Multi-Attention Fusion Network for Stroke Diagnosis}
       
\author{Aysen Degerli, Pekka Jäkälä, Juha Pajula, Milla Immonen, Miguel Bordallo López

\thanks{A. Degerli and J. Pajula are with the VTT Technical Research Centre of Finland, Tampere, Finland (e-mail:name.surname@vtt.fi).}
\thanks{P. Jäkälä is with the Kuopio University Hospital, Kuopio, Finland (e-mail: pekka.jakala@pshyvinvointialue.fi).}
\thanks{M. Immonen is with the Lapland University of Applied Sciences, Rovaniemi, Finland (e-mail: milla.immonen@lapinamk.fi).}
\thanks{M. B. López is with the University of Oulu, Oulu, Finland (e-mail: miguel.bordallo@oulu.fi).}
\thanks{This study was supported by Stroke-data project under Business Finland Grant 3617/31/2019.}}

\maketitle

\begin{abstract}
Stroke is a major cause of mortality and disability worldwide from which one in four people are in danger of incurring in their lifetime. The pre-hospital stroke assessment plays a vital role in identifying stroke patients accurately to accelerate further examination and treatment in hospitals. Accordingly, the National Institutes of Health Stroke Scale (NIHSS), Cincinnati Pre-hospital Stroke Scale (CPSS) and Face Arm Speed Time (F.A.S.T.) are globally known tests for stroke assessment. However, the validity of these tests is skeptical in the absence of neurologists and access to healthcare may be limited. Therefore, in this study, we propose a motion-aware and multi-attention fusion network (MAMAF-Net) that can detect stroke from multimodal examination videos. Contrary to other studies on stroke detection from video analysis, our study for the first time proposes an end-to-end solution from multiple video recordings of each subject with a dataset encapsulating stroke, transient ischemic attack (TIA), and healthy controls. The proposed MAMAF-Net consists of motion-aware modules to sense the mobility of patients, attention modules to fuse the multi-input video data, and 3D convolutional layers to perform diagnosis from the attention-based extracted features. Experimental results over the collected Stroke-data dataset show that the proposed MAMAF-Net achieves a successful detection of stroke with 93.62\% sensitivity and 95.33\% AUC score.
\end{abstract}

\begin{IEEEkeywords}
Stroke, Transient Ischemic Attack, NIHSS, Deep Learning, Self-attention.
\end{IEEEkeywords}

\section{Introduction}
\IEEEPARstart{S}{troke} is a cerebrovascular disease that World Health Organization (WHO) recognizes as the second leading cause of mortality worldwide. According to Tsao \textit{et al.} \cite{strokeStatistics}, stroke puts one to death in every $3,5$ minutes on average. Additionally, World Stroke Organization revealed that the number of stroke survivors has almost doubled over the years with an increase of younger patients having stroke \cite{feigin2022world}. Hence, stroke is no longer only a disease of the elderly and its global cost sharply increases. The severity of stroke varies from recovery to severe disability or even death \cite{gill2020recovery}. Consequently, its early and accurate diagnosis is vital for the prevention of mortality and disability. 

Cerebrovascular disease encompasses conditions of blood vessels supplying the brain \cite{andrade2012systematic}. Accordingly, stroke types are categorized as cerebrovascular accident (CVA) and transient ischemic attack (TIA). A cerebrovascular accident can either be ischemic stroke or hemorrhagic stroke. Ischemic stroke is a common type of CVA that occurs due to a blockage of blood supply to the brain, whereas hemorrhagic stroke is a more severe condition of CVA caused by bleeding into the brain due to rupture of a blood vessel \cite{unnithan2022hemorrhagic}. According to WHO, stroke develops symptoms lasting $24$ hours or longer (or leading to death). On the other hand, TIA is a mini-stroke, where its symptoms last less than $24$ hours \cite{world2005steps}. The literature reveals that TIA heralds the actual attack of a future ischemic stroke \cite{amarenco2020transient}. Therefore, an accurate diagnosis of TIA is crucial for the early detection of ischemic stroke.

The presence of a stroke can be recognized by imaging techniques, electroencephalography (EEG), and assessment tests based on physical examination. The imaging techniques have great significance in stroke prognosis, where abnormalities caused by a stroke in the brain and retinal arterioles can be revealed in Computed Tomography (CT) scans or Magnetic Resonance Imaging (MRI) \cite{birenbaum2011imaging} and retinal fundus images \cite{jeena2019retina}, respectively. Moreover, changes in the EEG patterns after a stroke almost immediately reveal brain ischemia \cite{doi:10.1161/STROKEAHA.120.030150}. Even though imaging tools and EEG recordings are sensitive to accurate stroke assessment, these techniques are costly and time-consuming \cite{kaur2022early}. Therefore, stroke assessment based on physical examination is promising for faster and cheaper diagnosis with pre-hospital or acute stroke assessment tests such as the National Institutes of Health Stroke Scale (NIHSS), Face Arm Speed Time (F.A.S.T.), and Cincinnati Pre-hospital Stroke Scale (CPSS). These tests try to assess stroke from any presence of unilateral facial droops, arm drifts, and speech disorders. However, stroke is highly heterogeneous for a straightforward  definition. In fact, Kasner \textit{et al.} \cite{doi:10.1161/01.STR.30.8.1534} and Meyer \textit{et al.} \cite{doi:10.1111/j.1747-4949.2009.00294.x} emphasize that NIHSS charts are merely approximate representations and prone to several issues such as time consumption, unreliability, and bias, primarily due to miscommunications that can occur during the testing process. Hence, technology is necessary to prevent subjective assessment \cite{warlow1998epidemiology}. 

Technological developments have improved machine learning (ML) to reach an outstanding performance in the healthcare domain with models achieving accurate, fast, and low-cost detection of various diseases and medical emergency conditions. Accordingly, ML is indeed promising for the standardization of stroke diagnosis. However, in the literature, only the studies \cite{yu2020toward} and \cite{lee2022deep} have tried to emulate F.A.S.T. and CPSS tests for automatic stroke detection by deep learning (DL) using examination videos. Additionally, the datasets for stroke detection are rather limited and the existing ones include only stroke and healthy controls which is unreliable for detecting patients having TIA. Furthermore, previous studies perform stroke analysis over the detected faces or facial \& body landmarks, where these models become prone to the errors of face and body detectors. In order to address the aforementioned limitations, in this study, we propose a novel motion-aware and multi-attention fusion network (MAMAF-Net) for stroke diagnosis using the multiple examination videos recorded from each subject as illustrated in Fig.~\ref{fig:framework}. We evaluated MAMAF-Net over the collected Stroke-data dataset that includes examination video recordings of $84$ stroke, $10$ TIA patients, and $54$ healthy controls. Accordingly, the key steps of our proposed MAMAF-Net are listed as follows:
\begin{itemize}
    \item MAMAF-Net consists of multi-input channels each inserted into motion-aware modules for extracting features representing the mobility capacity of subjects.
    \item The outputs of motion-aware modules are fused via attention-based fusion to support the dominant contributions of multi-input channels.
    \item The multi-attention fusion module is attached to $3$D convolutional layers to perform stroke diagnosis.
\end{itemize}

The rest of the paper is organized as follows. In Section \ref{sec:related_work}, a literature review including the prior works is provided. Then, the proposed methodology is explained in Section \ref{sec:methodology}, and details of data collection are given in Section \ref{sec:cohort}. Experimental results are reported in Section \ref{sec:results}, and technical \& clinical implications, limitations, and future work are discussed in Section \ref{sec:discussions}. Lastly, Section \ref{sec:conclusions} concludes the study.

\section{Related Work}\label{sec:related_work}
Recent improvements in machine learning and deep learning algorithms led to state-of-the-art performance in many computer vision tasks including image classification and segmentation, object detection, and emotion recognition. In healthcare, these algorithmic improvements are reflected in computer-aided diagnosis, where various diseases and medical emergency conditions are diagnosed with the support of ML models leading to outstanding performance. In the literature, computer-aided stroke diagnosis with deep learning-based approaches have been widely studied over CT scans, MRIs, fundus images, and EEG signals. Accordingly, Convolutional Neural Networks (CNNs) have been adapted to perform the classification of stroke in many studies using imaging techniques \cite{ZHU2022147}. Sheth \textit{et al.} \cite{doi:10.1161/STROKEAHA.119.026189} reported $90\%$ area under the curve (AUC) score for stroke detection over CT images with their proposed CNN model. A recent study in \cite{diagnostics12040807} has utilized transfer learning from state-of-the-art deep learning models using CT scans and achieved outstanding performance in stroke detection with $98\%$ AUC. Moreover, many studies \cite{zhang2021stroke, doi:10.1148/radiol.220882, 8969052, al2022deep} have tried similar approaches using CNNs and conventional machine learning algorithms with features extracted from pre-trained networks to detect stroke-related anomalies in MRIs. In fundus imaging, a similar path was followed in several studies \cite{jeena2021comparative, jcm11247408, jeena2019stroke}, where authors compared the stroke detection performance between the hand-crafted features and CNN models using transfer learning. Lastly, in EEG analysis, several studies \cite{s21134269, 9847883, doi:10.1063/5.0098733} reported that machine learning models achieved high performance for the diagnosis of stroke.

Although the aforementioned machine learning algorithms using biosignals and biomedical images are promising for computer-aided stroke diagnosis, acquiring bio-data is costly and time-consuming. Hence, stroke assessment tests are the first preference for the diagnosis in emergency conditions, where stroke symptoms such as facial droops, arm drifts, and speech disorders can be recognized and scored using the CPSS \cite{tarkanyi2020detailed} and NIHSS \cite{doi:10.1161/STROKEAHA.116.015434} tests, respectively. However, the utilization of these tests under the scarcity of neurologists and healthcare access prevents effective recognition and scoring of stroke in emergency situations. Therefore, there is a need for an automated test for the standardization of  stroke diagnosis, which follows a similar procedure as in the certified stroke assessment tests. Nevertheless, only the studies \cite{yu2020toward} and \cite{lee2022deep} have attempted to perform automatic recognition of stroke over examination videos using deep learning models. Yu \textit{et al.} \cite{yu2020toward} have tried to detect stroke with two-stream networks, where they use as inputs the frame sequences and audio of examination videos that are recorded while patients perform a set of vocal sets. Even though the two-streaming approach is promising, the body movements, which are an important indicator of stroke diagnosis, are missing from the analysis. On the other hand,  Lee \textit{et al.} \cite{lee2022deep} has included comprehensive video data for the analysis of body movements in addition to facial expressions. However, existing studies perform evaluations over limited categories having data only from stroke patients and healthy controls without data from TIA patients and they are bonded on the face \& body detector algorithms that bring several issues for clinical practice and use.

\section{Methodology}\label{sec:methodology}
In this section, we first describe the problem, which is stroke diagnosis using neurological examination videos. Then, we introduce the proposed deep learning model, MAMAF-Net that performs an end-to-end automatic stroke detection.

\subsection{Problem Definition}
In this study, we propose a novel model to detect any stroke-related assets in patients using neurological examination videos. The problem is formulated as a binary classification task, where the positive class is formed by stroke and TIA patients, and the negative class includes only healthy controls. The examination videos are recorded by the nurse during the NIHSS assessment at the Stroke Unit of the hospital. Accordingly, four video recordings are selected for the analysis that corresponds to facial palsy, best gaze, best language, and motor arm items in the NIHSS list.

\begin{figure*}[t!]
    \centering
    \includegraphics[width=.85\linewidth]{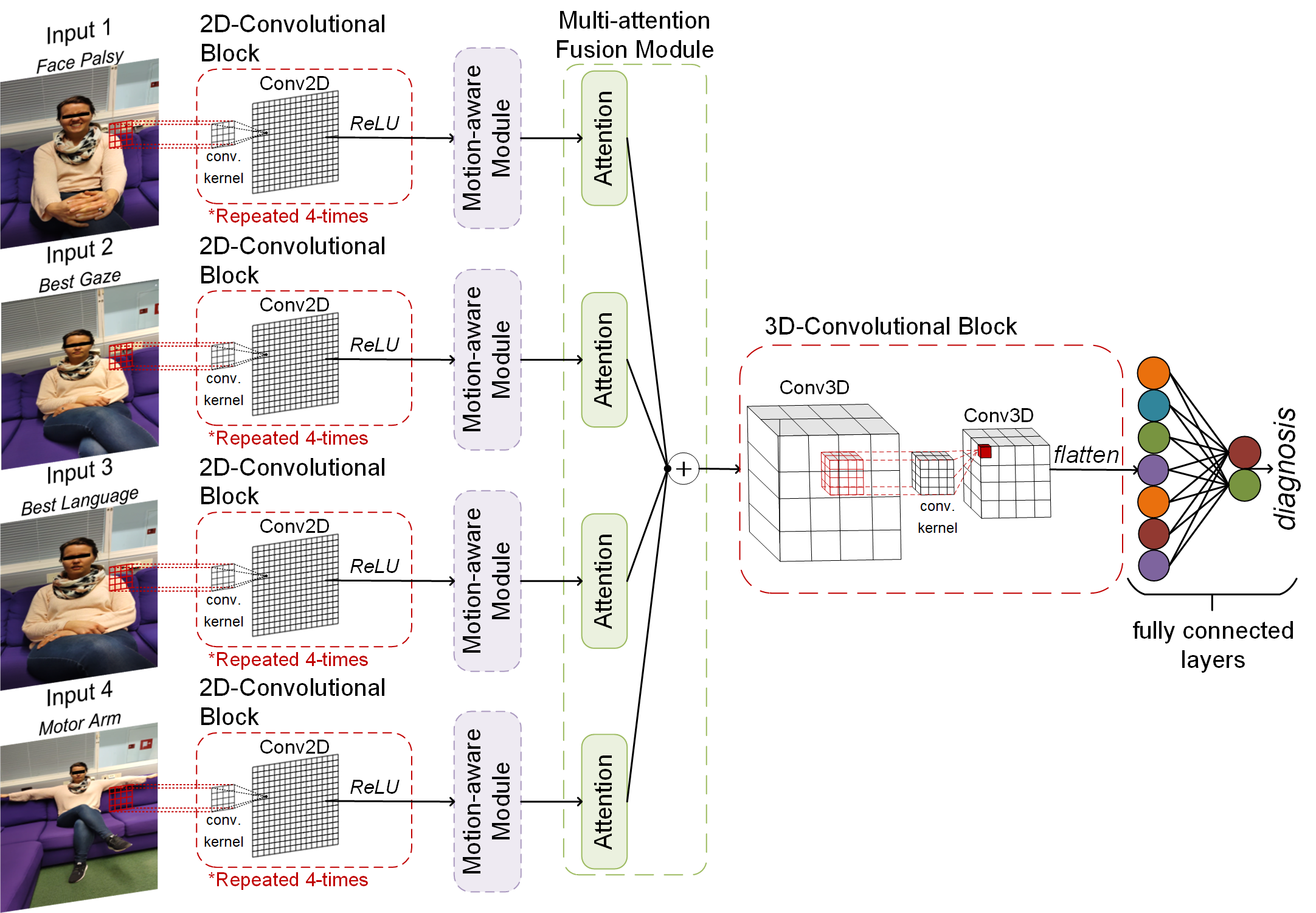}
    \caption{The proposed MAMAF-Net architecture is illustrated.}
    \label{fig:framework}
\end{figure*}

Let us denote a video recording as $\mathbf{X}\in \mathbb{R} ^{N \times w \times h \times c}$, where $N$ is the sequence length, $w$ is width, $h$ is height, and $c$ is the color channel of video frames. The dataset including the video recordings $\mathbf{I} = \{\mathbf{X}_1, \mathbf{X}_2, \mathbf{X}_3, \mathbf{X}_4\}$ and their corresponding ground-truth labels $\mathbf{y}\in \{0, 1\}$ is denoted as $D={\{\mathbf{I}_s, \mathbf{y}_s\}}_{s=1}^S$, where $S$ is the total number of samples and $s$ is a sample in the dataset. Accordingly, the proposed MAMAF-Net maps input $\mathbf{I}$ to predicted class, $\mathbf{\hat{y}}: \mathbf{\hat{y}} \leftarrow{} \Theta_{\mu, \alpha, \kappa}(\mathbf{I}, \mathbf{y})$, where the network $\Theta$ consists of motion-aware modules $\mu$, multi-attention fusion module $\alpha$, and convolutional blocks $\kappa$ as illustrated in Fig.~\ref{fig:framework}.

\subsection{MAMAF-Net: Motion-aware and Multi-attention Fusion Network}
The key components of the proposed MAMAF-Net are described in this section. The proposed network consists of multi-input channels each attached to a $2$D-convolutional block following a motion-aware module. Then, the branches are connected through the multi-attention fusion module, and $3$D-convolutional blocks to reduce the dimension prior to fully connected layers as it is depicted in Fig.~\ref{fig:framework}.

\textbf{$2$D-Convolutional Block.} The network $\Theta$ consists of four identical $2$D-convolutional blocks, where each block is composed of $2$D convolutional layers and  Rectified Linear Unit (ReLU) activation functions. A single $2$D-convolutional block, $\kappa \in \{b_l, \omega_l\}_{l=1}^L$ consists of $L=4$ number of convolutional layers each attached to a ReLU activation function, where $b$ and $\omega$ are bias term and weights of the convolutional layer. For each convolutional layer, a kernel size of $k=(3\times3)$ and a stride of $\delta=(2\times2)$ are used in order with the number of filters $f=\{64, 32, 16, 8\}$, where padding is applied with zeros.

\begin{figure*}[t!]
    \centering
    \includegraphics[width=.8\linewidth]{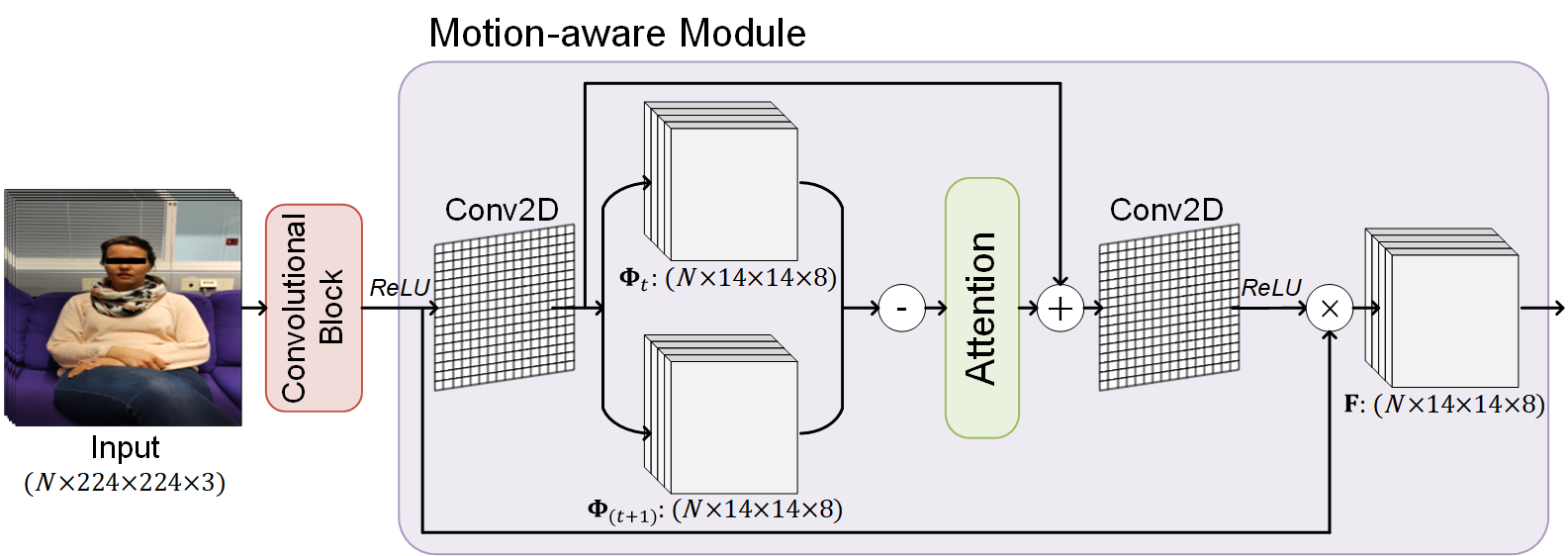}
    \caption{A branch of the network, where the motion-aware module is depicted in detail.}
    \label{fig:motion_module}
\end{figure*}

\textbf{Motion-aware Module.} The output of each $2$D-convolutional block is attached to a motion-aware module, where features $\mathbf{F} \leftarrow{} \mu_{\Theta}(.)$ representing the mobility are extracted as illustrated in Fig.~\ref{fig:motion_module}. In our study, we are inspired by \cite{gebotys2022m2a}, where they have proposed a motion-aware attention (M2A) mechanism to take advantage of inherent motion in video recordings. In order to simply adapt M2A to our task, we have modified their mechanism. Accordingly, in our motion-aware module, a $2$D convolutional layer with a number of filters $f=8$ and a kernel size of $(3\times3)$ is connected to the $2$D convolutional block. The output of the convolutional layer is denoted as $\mathbf{\Phi}_t \in \mathbb{R}^{N \times 14 \times 14 \times 8}$, where $t$ is the representation of the current time frame. Then, $\mathbf{\Phi}_t$ is shifted with a rolling operation to form $\mathbf{\Phi}_{(t+1)}$ by simply copying the first frame of $\mathbf{\Phi}_t$ to the last frame of $\mathbf{\Phi}_{(t+1)}$. The motion information is extracted by computing the difference $(\mathbf{\Phi}_t - \mathbf{\Phi}_{(t+1)})$, which is then followed by a dot-scaled attention layer to dominate the relevant motion patterns. To further enhance motion patterns, $\mathbf{\Phi}_t$ is added to the output of the attention layer. Then, a $2$D convolutional layer with a number of filters $f=8$ and a kernel size of $(3\times3)$ followed by a ReLU activation function is applied. Lastly, the feature representation $\mathbf{F} \in \mathbb{R}^{N \times 14 \times 14 \times 8}$ is formed by the element-wise multiplication of the input of the motion-aware module and the output of the last convolutional layer.

\begin{figure*}[b!]
    \centering
    \includegraphics[width=.8\linewidth]{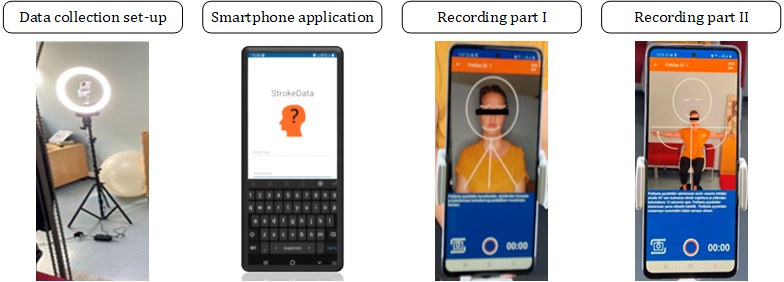}
    \caption{The data collection process using the Stroke-data smartphone application.}
    \label{fig:data_collection_app}
\end{figure*}

\textbf{Multi-attention Fusion Module.} The multi-input channels $\mathbf{X} \in \mathbb{R}^{N \times 224 \times 224 \times 3}$ are mapped to the motion features $\mathbf{F} \in \mathbb{R}^{N \times 14 \times 14 \times 8}$, which are the inputs of the multi-attention fusion module $\alpha$ attached to attention layers. The dot-scaled attention layer, which is proposed by \cite{vaswani2017attention} is used in our proposed network. The attention function maps a query $Q$, a set of key $K$ and value $V$ pairs to an output, which is the weighted sum of the values. The weights on the values are computed as follows:
\begin{equation}
    \text{Attention}(Q, K, V)= \text{softmax}\left(\frac{QK^T}{\sqrt{d_k}}\right)V,
\end{equation}
\noindent where $\text{softmax}$ is a function, and $d_k$ is the dimension of queries and keys that is used for the scaling factor. Lastly, the outputs of the attention layers $\mathbf{A} = \{\mathbf{A}_1, \mathbf{A}_2, \mathbf{A}_3, \mathbf{A}_4\} \in \mathbb{R}^{N \times 14 \times 14 \times 8}$ are added to fuse the multiple features extracted from multi-input channels corresponding to $\{\mathbf{X}_1, \mathbf{X}_2, \mathbf{X}_3, \mathbf{X}_4\}$. 

\textbf{$3$D-Convolutional Block.} The last component of the proposed network that is attached to the output of multi-attention fusion module $\mathbf{A}$ is composed of $3$D convolutional layers to reduce the dimension across the channels simultaneously. Accordingly, the block consists of two $3$D convolutional layers with a kernel size of $k=(3 \times 3 \times 3)$, $f=3$ number of filters, and strides of $(5 \times 2 \times 2) \text{ and } (5 \times 1 \times 1)$ followed by ReLU activation functions. At last, its output $\mathbf{\Psi} \in \mathbb{R}^{\frac{N}{25} \times 7 \times 7 \times 3}$ is vectorized and attached to fully connected layers, where the output layer consists of $2$-neurons and the softmax activation function for the diagnosis.

In this study, in comparison to the proposed MAMAF-Net model, the state-of-the-art networks: ResNet50 \cite{he2016deep} and DenseNet-121 \cite{huang2017densely}, where their weights are initialized with the ImageNet dataset weights by transfer learning are used as the baseline methods. Accordingly, we extract deep features from each frame of input video sequences using the state-of-the-art deep networks, where the extracted deep features $\mathbf{D} \in \mathbb{R}^{N \times 1000}$ are fused by addition layer. Then, the fused feature data $\mathbf{D}$ is flattened and attached to fully connected layers, where the output layer consists of $2$-neurons and the softmax activation function for the diagnosis. Alternatively, we embed the multi-attention fusion module to state-of-the-art models to form MAF-ResNet50 and MAF-DenseNet-121 models to investigate the performance of our proposed module in the diagnosis.

\begin{table*}[t!]
\centering
\caption{The characteristics of the study cohort, where $n$ indicates the number of subjects.}
\resizebox{.9\textwidth}{!}{
\begin{tabular}{llcccccc}
& & \multicolumn{2}{c}{Control} & \multicolumn{2}{c}{Stroke} & \multicolumn{2}{c}{TIA}
\\ \cmidrule(lr){3-4}\cmidrule(lr){5-6}\cmidrule(lr){7-8}
             &          & \begin{tabular}[c]{@{}c@{}}Female \\ $(n=32)$ \end{tabular}  & \begin{tabular}[c]{@{}c@{}}Male \\ $(n=22)$ \end{tabular}  & \begin{tabular}[c]{@{}c@{}}Female \\ $(n=35)$ \end{tabular}  & \begin{tabular}[c]{@{}c@{}}Male \\ $(n=49)$ \end{tabular}  &  \begin{tabular}[c]{@{}c@{}}Female \\ $(n=4)$ \end{tabular} &  \begin{tabular}[c]{@{}c@{}}Male \\ $(n=6)$ \end{tabular} \\ \toprule
Age \\(years)  
             & Mean     & $65$    & $66$     & $65$     & $68$     & $64$     & $74$     \\
             & Maximum  & $78$    & $83$     & $83$     & $87$     & $79$     & $85$     \\
             & Minimum  & $29$    & $36$     & $31$     & $44$     & $46$     & $57$     \\
             &          &         &          &          &          &          &          \\
Weight \\(kg)  & Mean     & $71.63$ & $80.82$  & $77.69$  & $82.56$  & $73$     & $84.67$  \\
             & Maximum  & $130$   & $95$     & $148$    & $140$    & $75$     & $95$     \\
             & Minimum  & $41$    & $65$     & $42$     & $56$     & $72$     & $70$     \\
             &          &         &          &          &          &          &          \\
Height \\(cm) & Mean     & $162.66$ & $176.91$ & $163.20$ & $176.63$ & $165.23$ & $175.25$ \\
            & Maximum  & $178$    & $182$    & $176$    & $190$    & $168.9$  & $190$    \\
            & Minimum  & $150$    & $172$    & $150$    & $165$    & $162$    & $168$    \\
            &          &          &          &          &          &          &           \\
Education level \\(number) & Primary education & $12$& $10$     & $12$     &                                $17$     & $2$ & $3$\\
                            & Upper secondary education & $12$& $7$      & $18$     & $20$     & $-$ & $3$\\ 
                            & Bachelors or equivalent & $8$ & $3$  & $3$      & $5$      & $1$ & $-$ \\
                           & Masters/Doctoral or equivalent & $-$ & $2$      & $2$      & $7$      & $1$ & $-$ \\
                           &   &     &     &          &          &     &    \\
Work Status \\(number)    & Employee & $7$ & $5$      & $10$     & $10$     & $2$ & $1$\\
                         & Entrepreneur    & $1$ & $2$      & $3$      & $7$      & $-$ & $-$\\
                         & Retired & $23$& $15$     & $21$     & $32$     & $2$ & $5$\\
                         & Student  & $1$ & $-$      & $-$      & $-$      & $-$ & $-$\\
                         & Disable to work    & $-$ & $-$      & $1$      & $-$      & $-$ & $-$ \\ \bottomrule
\end{tabular}}
\label{tab:characteristics}
\end{table*}

\section{Cohort}\label{sec:cohort}
The clinical dataset for this study was collected at Kuopio University Hospital and Oulu University Hospital between the years $2021$ and $2022$ under the IRB-approved Stroke-data project. The data was recorded at the Stroke Units of the hospitals during stroke assessments of patients with the NIHSS protocol using the dedicated mobile application. Each patient’s performance in the NIHSS protocol was evaluated by study nurses, who had received the NIHSS protocol certification. The diagnosis of stroke and TIA were done by neurologists at the hospital during the patient's stay at the hospital through clinical evidence and other sensing and imaging modalities, i.e., EEG and MRI. Accordingly, the cohort consists of $148$ participants from stroke and TIA patients as well as healthy controls. Table~\ref{tab:characteristics} shows the characteristics of the cohort, where in total $71$ females and $77$ males are included in the study. It can be observed from the table that the participants are mostly retired and the average age varies from $64$ to $74$ through stroke \& TIA patients and healthy controls.

For the collection of the video data, we have developed a smartphone application called the Stroke-data application, which is used to record videos of the patients during the NIHSS protocol as shown in Fig.~\ref{fig:data_collection_app}. The collection is performed with a tripod stand and a circular light source, where the smartphone with the Stroke-data application is placed in front of the patient while each NIHSS task is recorded separately. The collection procedure takes approximately $10$ minutes, and only the research nurse uses the application intended for the video data recording. In the measurement protocol, the smartphone is placed in such a way that only the patient under examination is visible in the camera. In the initial phase, the target is specifically the face, in which case the smartphone is placed as close to the patient as possible. In the last phase, the patient's arms must be completely visible on the screen as moving the arms; hence, the smartphone has to be moved further away from the patient. Then, the task starts, and the research nurse receives step-by-step instructions from the application. The NIHSS steps performed for this study are as follows: i) \textit{facial palsy}: the patient is asked to smile, show teeth and raise eyebrows in succession a few times, ii) \textit{best gaze}: the patient is asked to follow an object left and right with his/her eyes while keeping his/her head still to check the strength at extraocular muscles, iii) \textit{best language}: the patient is asked to read aloud the sentences presented on paper, and iv) \textit{motor arm}: the patient is asked consecutively to lift his/her left arm, right arm, and both arms to the sides at a $90^{\circ}$ angle with eyes closed. The study nurses recorded the NIHSS scores for each task during the data recording session according to the NIHSS protocol.

\section{Results}\label{sec:results}
In this section, we present the experimental setup and report the results over the Stroke-data dataset.

\begin{table*}[t!]
  \centering 
  \caption{Average detection results (\%) of MAMAF-Net and state-of-the-art deep models with respect to the sequence length $N$ evaluated over $5$-fold CV scheme, where the highest scores are highlighted in bold.}
  \resizebox{.92\linewidth}{!}{
  \begin{tabular}{clcccccc}
  \toprule
  \textbf{Sequence Length} & \textbf{Method} & \textbf{Sensitivity} & \textbf{Specificity} & \textbf{Precision} & \textbf{F1-Score} & \textbf{Accuracy} & \textbf{AUC} \\
  \midrule
  \multirow{5}{*}{$(N = 25)$} & MAMAF-Net & $87.23$	& $83.33$ & $90.11$ & $88.65$ & $85.81$ & $92.93$\\ 
  & MAF-ResNet50 & $85.11$ & $68.52$ & $82.47$ & $83.77$ & $79.05$ & $86.70$ \\ 
  & MAF-DenseNet-121 & $92.55$ & $70.37$ & $84.47$ & $88.32$ & $84.46$ & $89.99$ \\ 
  & ResNet50 & $84.04$ & $68.52$ & $82.29$ & $83.16$ & $78.38$ & $85.95$ \\ 
  & DenseNet-121 & $91.49$ & $74.07$ & $86.00$ & $88.66$ & $85.14$ & $92.65$ \\
  \midrule
  \multirow{5}{*}{$(N = 50)$} & MAMAF-Net & $84.04$ & $\textbf{88.89}$ & $\textbf{92.94}$ &	$88.27$ & $85.81$ & $93.60$ \\ 
  & MAF-ResNet50 & $81.91$ & $68.52$ & $81.91$ & $81.91$ & $77.03$  & $85.36$ \\ 
  & MAF-DenseNet-121 & $86.17$ & $77.78$ & $87.10$ & $86.63$ & $83.11$ & $93.12$ \\ 
  & ResNet50 & $81.91$ & $64.81$ & $80.21$ & $81.05$ & $75.68$ & $82.76$ \\
  & DenseNet-121 & $86.17$ & $75.93$ & $86.17$ & $86.17$ & $82.43$ & $91.86$ \\
  \midrule
  \multirow{5}{*}{$(N = 75)$} & MAMAF-Net & $\textbf{93.62}$ & $79.63$ & $88.89$ & $\textbf{91.19}$ & $\textbf{88.51}$ & $\textbf{95.33}$ \\
  & MAF-ResNet50 & $81.91$ & $66.67$ & $81.05$ & $81.48$ & $76.35$ & $83.41$ \\ 
  & MAF-DenseNet-121 & $88.30$ & $72.22$ & $84.69$ & $86.46$ & $82.43$ & $91.41$ \\ 
  & ResNet50 & $81.91$ & $68.52$ & $81.91$ & $81.91$ & $77.03$ & $85.50$ \\
  & DenseNet-121 & $89.36$ & $74.07$ & $85.71$ & $87.50$ & $83.78$ & $90.03$ \\
  \midrule
  \bottomrule
  \end{tabular}}
  \label{tab:experiment} 
\end{table*}

\subsection{Evaluation Approach}
The experimental evaluations are performed in a stratified $5$-fold cross-validation (CV) scheme with a ratio of $80\%$ \textbf{training set} to $20\%$ \textbf{test set}. In the training phase, $20\%$ of the training set is separated for the \textbf{validation set} in each $5$-fold. Accordingly, the separation of the train, test, and validation sets is performed patient-wise. The standard performance metrics are computed over a cumulative confusion matrix as follows: \textit{sensitivity} is the rate of correctly identified stroke samples in the positive class, \textit{specificity} is the ratio of accurately detected control group samples among the negative class samples, \textit{precision} is the rate of correctly identified stroke patients among the samples predicted as a positive class, \textit{F1-Score} is the harmonic average between sensitivity and precision, \textit{accuracy} is the rate of correctly detected samples in the dataset. Additionally, the Receiver Operating Characteristics (ROC) curve is computed from which the area under the curve (AUC) scores are calculated.

The Stroke-data dataset is used for the evaluations, which includes a total of $148$ subjects with the ground truths of $54$ healthy controls and $94$ stroke \& TIA patients. The video recordings in the dataset are resized to $224 \times 224$ pixels to have compatible input dimensions for many state-of-the-art deep networks. The video sequence length $N$ is set to different values of $\{25, 50, 75\}$ for experimental investigations by taking equally distanced $N$ frames from a video recording. Additionally, we have investigated the impact of data augmentation in the training phase in order to overcome the issue of imbalanced class samples, where the video sequences $\mathbf{X} \in \mathbb{R}^{25 \times 224 \times 224 \times 3}$ are augmented up to $100$ samples for each class by randomly rotating the sequences at $90^{\circ}$, $180^{\circ}$, and $270^{\circ}$ angles, randomly flipping vertically \& horizontally, and the combination of these operations.

The networks are implemented with the TensorFlow library using Python and trained on an NVidia® Amper A$100$ GPU card with $40$ GB memory. Accordingly, the proposed MAMAF-Net is trained over $300$ epochs with a learning rate of $10^{-5}$ using categorical cross-entropy loss. The optimizer choice for the model is Adam with its default parameter settings. Lastly, in the training phase, the weights of the model, which has the minimum loss over the validation set are used for the testing. For the baseline models, the same configurations are valid except for a learning rate of $10^{-3}$ is used for better convergence.

\begin{figure}[b!]
         \centering
         \begin{subfigure}[t]{.49\linewidth}
         {\includegraphics[width=\linewidth]{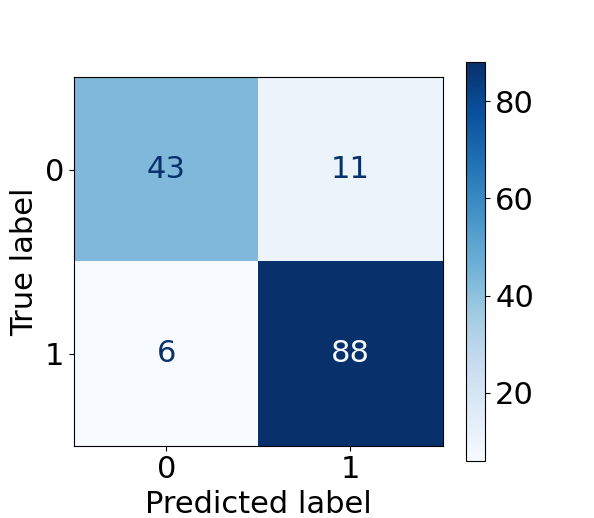}}
         \label{fig:mamafCM}
         \caption{}
         \end{subfigure}
    \hfill
         \centering
         \begin{subfigure}[t]{.49\linewidth}
         {\includegraphics[width=\linewidth]{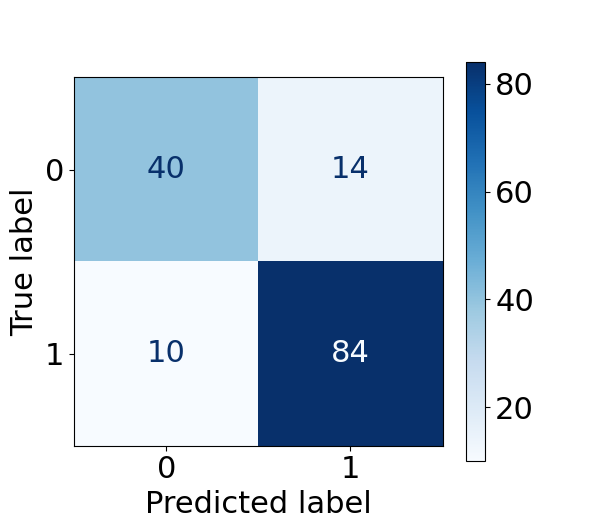}}
         \label{fig:densenetCM}
         \caption{}
         \end{subfigure}         
     \centering
     \caption{Confusion matrices of the MAMAF-Net (a) and DenseNet-121 (b) models with sequence length $(N=75)$.}
     \label{fig:confusion_matrices}
\end{figure}

\begin{figure*}[t!]
    \centering
    \includegraphics[width=.78\linewidth]{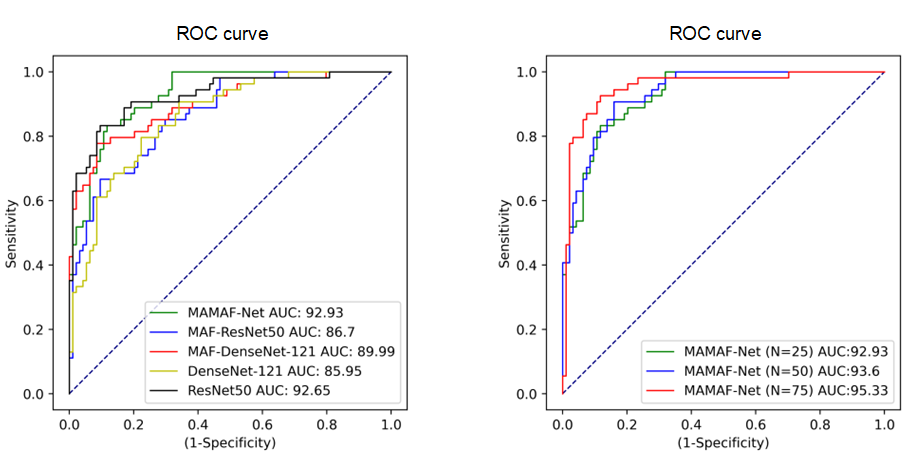}
    \caption{The ROC curves, where the analysis of AUC is computed for each model with video sequence length $(N=25)$ at the left side, and the impact of the sequence length of the proposed MAMAF-Net model is plotted at the right side.}
    \label{fig:ROC_curve}
\end{figure*}

\subsection{Experimental Results} 
In this section, the experimental results of the test (unseen) set are given for the MAMAF-Net model. Table~\ref{tab:experiment} presents the average detection results of the proposed MAMAF-Net and state-of-the-art computed over the $5$-folds. Accordingly, in the table, the video sequence length is increased to investigate the performance of MAMAF-Net model for stroke diagnosis. It can be seen that for each video sequence, the highest AUC scores are obtained by the proposed MAMAF-Net model. On the other hand, among state-of-the-art deep models, the highest AUC score is achieved by DenseNet-121 with a $92.65\%$ for the sequence length of $(N=25)$, where the performance is the closest by $0.28\%$ to the proposed MAMAF-Net with $92.93\%$ AUC. Additionally, the proposed MAMAF-Net has achieved a successful diagnosis performance with a sensitivity level of $93.62\%$ and F$1$-Score of $91.19\%$ for the largest sequence length.

\begin{table*}[b!]
  \centering 
  \caption{Average detection results (\%) of MAMAF-Net and state-of-the-art deep models with data augmentation in the training phase evaluated over $5$-fold CV scheme, where the sequence length is $(N=25)$ and the highest scores are highlighted in bold.}
  \resizebox{.78\linewidth}{!}{
  \begin{tabular}{lcccccc}
  \toprule
   \textbf{Method} & \textbf{Sensitivity} & \textbf{Specificity} & \textbf{Precision} & \textbf{F1-Score} & \textbf{Accuracy} & \textbf{AUC} \\
   \midrule
   MAMAF-Net & $89.36$ & $\textbf{87.04}$ & $\textbf{92.31}$ & $\textbf{90.81}$ & $\textbf{88.51}$ &  $\textbf{95.23}$ \\ 
   MAF-ResNet50 & $86.17$ & $72.22$ & $84.38$ & $85.26$ & $81.08$ & $84.57$ \\ 
   MAF-DenseNet-121 & $\textbf{92.55}$ & $72.22$ & $85.29$ & $88.78$ & $85.14$ & $91.55$ \\ 
   ResNet50 & $81.91$ & $74.07$ & $84.62$ & $83.24$ & $79.05$ & $84.54$ \\
   DenseNet-121 & $\textbf{92.55}$ & $75.93$ & $87.00$ & $89.69$ & $86.49$ & $92.65$ \\
   \midrule
   \bottomrule
  \end{tabular}}
  \label{tab:experiment_aug} 
\end{table*}

Further analysis reveals that increasing the sequence length improves the performance of the proposed model in terms of the AUC score as it can be depicted in Fig.~\ref{fig:ROC_curve}. Accordingly, the confusion matrices of the MAMAF-Net and DenseNet-121 models for sequence length $(N=75)$ are given in Fig.~\ref{fig:confusion_matrices}, where their sensitivity levels are $93.62\%$ and $89.36\%$, respectively. The confusion matrices reveal that the proposed model can correctly detect $88$ stroke patients, and $43$ healthy controls, whereas the state-of-the-art DenseNet-121 model detects $84$ stroke patients and $40$ healthy controls accurately. Moreover, we generally see an improvement in sensitivity metric in state-of-the-art models by utilizing the multi-attention fusion in the model, especially for the smallest sequence length of $(N=25)$. Accordingly, the ROC curve plots of the models as the sequence length is $(N=25)$ are given in Fig.~\ref{fig:ROC_curve}, where each model has achieved an AUC score of $>85\%$ for stroke diagnosis. We emphasized the results of $(N=25)$ in Fig.~\ref{fig:ROC_curve} since it can generally achieve comparable results to the leading sequence length of $(N=75)$ with less computational complexity, especially, when considering an application of the proposed approach in devices with limited memory and battery.

\begin{figure*}[t!]
         \centering
         \begin{subfigure}[t]{0.48\textwidth}
         {\includegraphics[width=\textwidth]{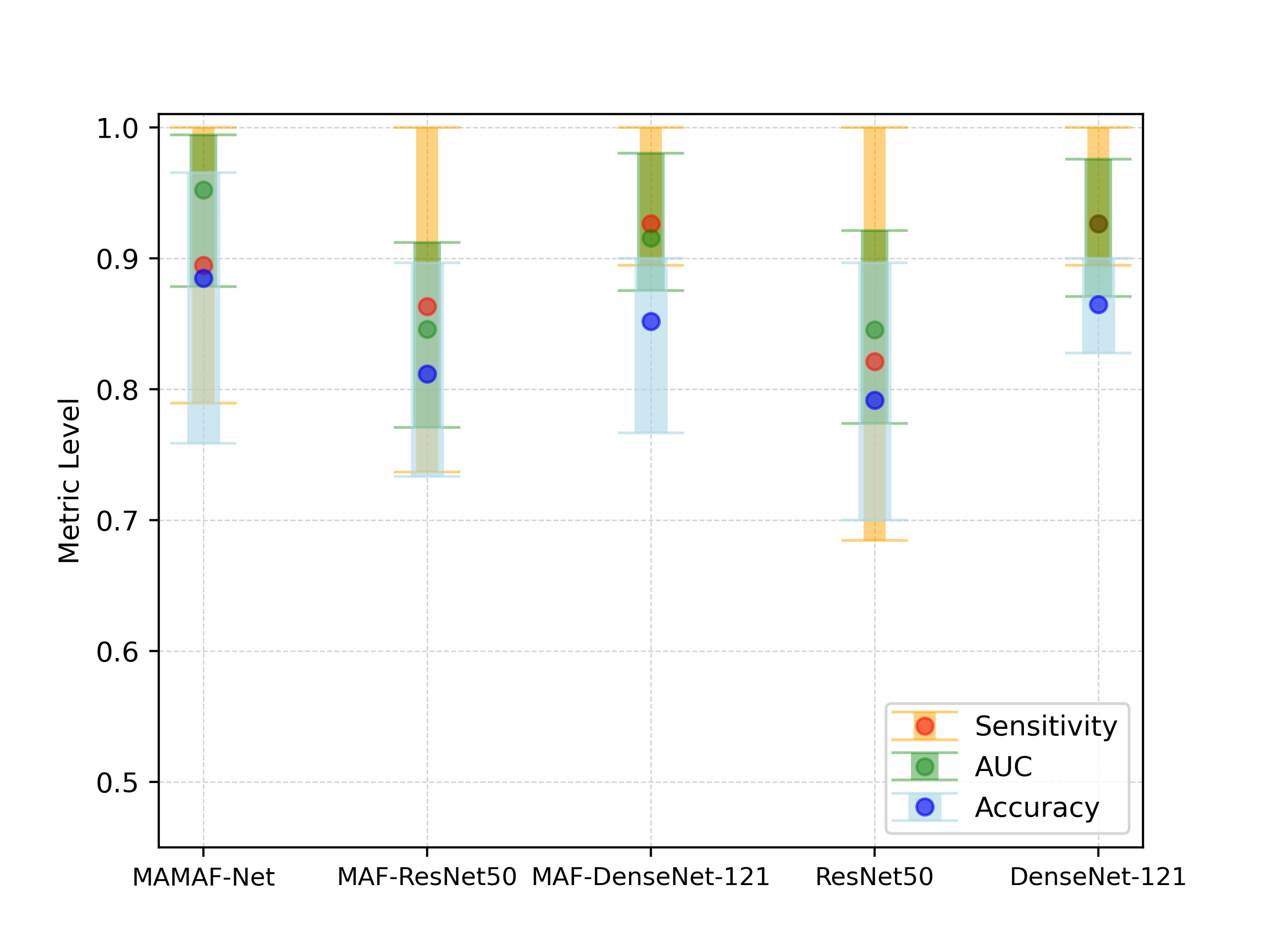}}
         \label{fig:error_bar_a}
         \caption{}
         \end{subfigure}
    \hfill
         \centering
         \begin{subfigure}[t]{0.48\textwidth}
         {\includegraphics[width=\textwidth]{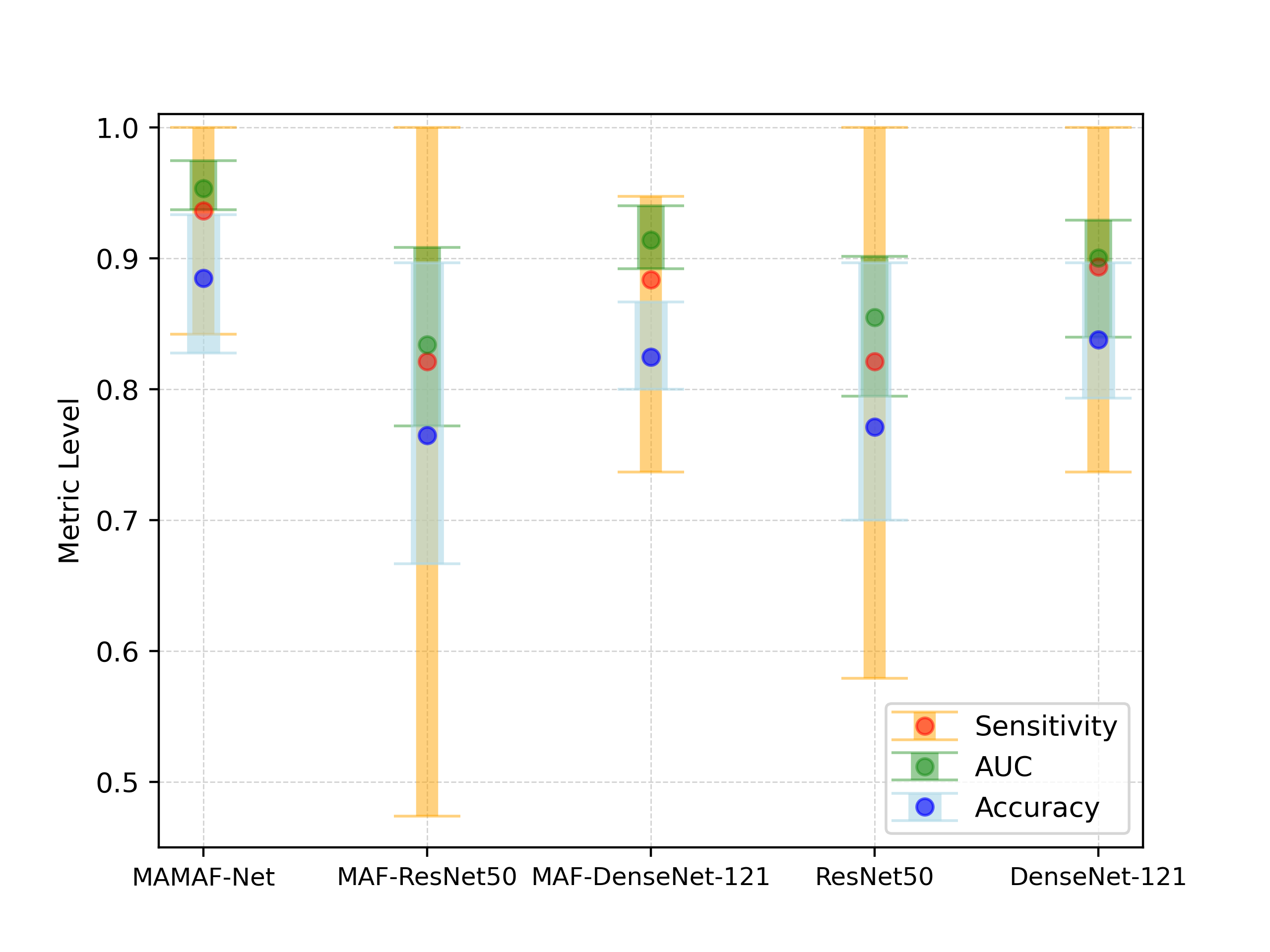}}
         \label{fig:error_bar_b}
         \caption{}
         \end{subfigure}         
     \centering
     \caption{The performance of each model is given for comparison with respect to sensitivity, accuracy, and AUC metrics, where the network configurations are indicated for the sequence length of a) $(N=25)$ with data augmentation, and b) $(N=75)$ without data augmentation.}
     \label{fig:error_bar}
\end{figure*}

\begin{figure*}[b!]
         \centering
         \begin{subfigure}[t]{0.48\textwidth}
         {\includegraphics[width=\textwidth]{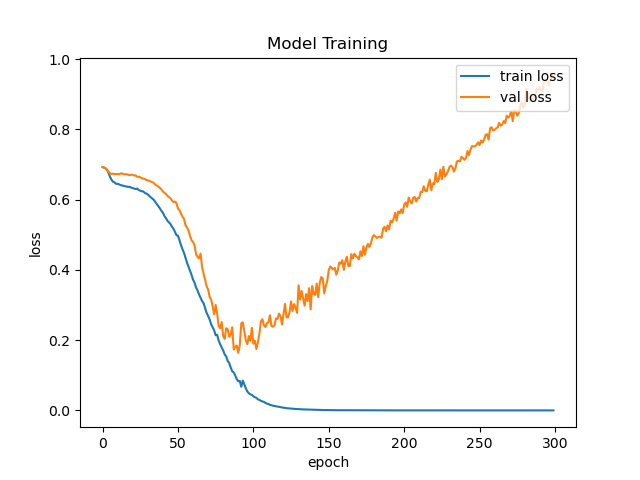}}
         \label{fig:training_a}
         \caption{}
         \end{subfigure}
        \hfill
        \centering
         \begin{subfigure}[t]{0.48\textwidth}
         {\includegraphics[width=\textwidth]{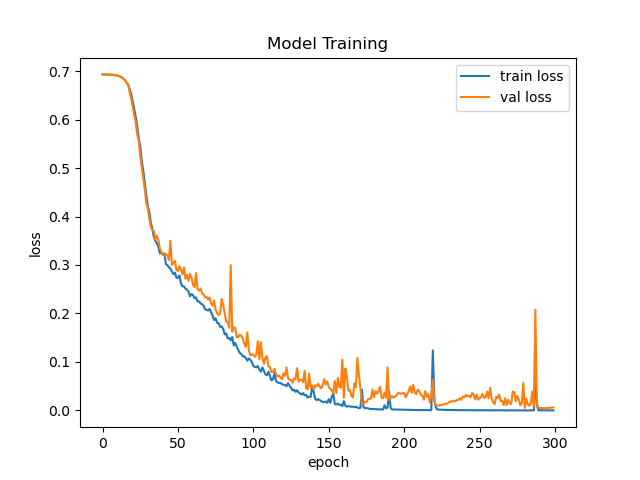}}
         \label{fig:training_b}
         \caption{}
         \end{subfigure}
     \centering
     \caption{The model training curves of the proposed MAMAF-Net are given with respect to train and validation loss, where the model is trained a) without data augmentation and b) with data augmentation over $300$ epochs.}
     \label{fig:training}
\end{figure*}

In this study, we further investigate the impact of data augmentation on the performance of the models as data augmentation is applied in the training phase. Table~\ref{tab:experiment_aug} shows the average detection results of the proposed MAMAF-Net and state-of-the-art deep models computed over $5$-folds. The analysis is performed for the smallest sequence length of $(N=25)$. Accordingly, the data augmentation helps to improve the performance of the proposed MAMAF-Net model with a boost of $2.13\%$ in sensitivity and $2.30\%$ in AUC score as compared to the performance of the model without data augmentation. The loss curves obtained during model training are given in Fig.~\ref{fig:training}, which reveals that the data augmentation helps the model to achieve a better convergence. Lastly, the performance of each model is given in Fig.~\ref{fig:error_bar}, which indicates the average, the highest, and the lowest levels of sensitivity, accuracy, and AUC metrics that are calculated over $5$-folds. Accordingly, Fig.~\ref{fig:error_bar} reveals that data augmentation in model training generally helps for the stabilization by reducing the standard deviation of performance among each folds. Additionally, MAMAF-Net achieves the highest performance for the sequence length $(N=75)$ as attaining a low standard deviation of $0.0159$, $0.0515$, and $0.0356$ for the metrics AUC, sensitivity, and accuracy, respectively. 

\section{Discussion}\label{sec:discussions}
In this study, we propose MAMAF-Net for the standardization of pre-hospital stroke assessment tests conducted at the emergency departments of hospitals. For this purpose, we compiled a dataset called Stroke-data that consists of $148$ subjects each including video recordings of the NIHSS steps, where face palsy, best gaze, best language, and motor arm steps are used as the multi-input channels to MAMAF-Net deep model. In the proposed network, we use the motion-aware module to extract features representing the motion patterns across a video sequence, and the multi-attention fusion module to fuse the extracted features as supporting the dominant contributions by dot-scaled attention layers. At the last block of the proposed network, $3$D convolutional layers are used for learning the patterns across dimensions while reducing the size of the tensor. 

\textbf{Technical implications.}
Contrary to previous studies, our compiled Stroke-data dataset is unique in terms of including TIA patients in addition to stroke and data from two distinct hospitals. This way, Stroke-data helps the proposed machine learning model to achieve a better and more reliable generalization of stroke for diagnosis. The computational complexity of the proposed model grows as the sequence length increases which also improves the stroke diagnosis performance. However, experimental results show that data augmentation can be the solution for this issue, where we reported that data augmentation can improve the performance of the model. In this way, video sequences with a smaller length can achieve a similar AUC score as a larger sequence length. Additionally, our study proposes for the first time an end-to-end solution for stroke diagnosis using several neurological examination videos recorded from the patients during NIHSS test. Hence, we show that machine learning algorithms are promising tools for the standardization of stroke assessment tests.

\textbf{Clinical implications.}
In clinical practice, the proposed model can be used in a smartphone application as an assistive tool in support of the pre-hospital stroke assessment, where the application would enable common emergency staff to perform NIHSS with minimal human bias and without additional training. Moreover, the application saves the video recordings in an encrypted and protected partition of the smartphone, which complies with data privacy regulations. 

\textbf{Limitations and Future Work.}
In computer-aided diagnostics, demographic representativeness is highly significant to ensure machine learning models are equitable and unbiased toward specific populations. Hence, it would be very beneficial to test our proposed model on open-access datasets, where data is gathered from diverse regions worldwide showing variations in the demographics. Unfortunately, at this stage, there are no publicly available video datasets for stroke detection. Additionally, due to the lack of TIA patients in the dataset, the proposed model is trained as a binary classification task, where the positive class is formed by stroke and TIA patients together. In future work, we plan to increase TIA patients in the dataset, so that the proposed model can be trained to discriminate stroke, TIA, and healthy controls as a multi-class problem. The ultimate problem is to regress the values of the NIHSS globally (as a total score) and also partially (as a sum of the partial scores) using a dataset, where the NIHSS protocol is fully completed. 

\section{Conclusions}\label{sec:conclusions}
This study contributes to the standardization of pre-hospital stroke assessment tests with the proposed machine learning-based approach. The current state-of-the-art that tackles the automatic detection of stroke using examination videos is rather limited. Hence, we propose a motion-aware and multi-attention fusion network, namely MAMAF-Net to overcome the existing issues in stroke diagnosis. In healthcare, achieving a reliable diagnosis with high sensitivity is crucial for patient treatment. Hence, in this study, in addition to stroke patients, we include TIA patients in the training and evaluation of MAMAF-Net to prevent missing any possible future ischemic strokes. Moreover, our proposed network is not dependent on any errors that may arise from face or facial \& body landmarks detection algorithms. Experimental results show that the proposed MAMAF-Net achieves the highest AUC with $95.33\%$ score among state-of-the-art deep classifiers. Consequently, the proposed MAMAF-Net can be deployed to a smartphone application, where stroke assessment can be performed in the absence of neurologists in emergency situations. Moreover, the proposed approach can be used in a variety of healthcare problems such as epileptic seizures and Parkinson’s disease diagnoses from video recordings.

\section*{Acknowledgments}
Authors would like to thank the study nurses Riitta Laitala, Saara Haatanen, Matti Pasanen, and Tanja Kumpulainen, research assistant Jari Paunonen, and senior scientist Timo Urhemaa for their contributions to data collection.

\bibliographystyle{IEEEtran}
\bibliography{IEEEtran}
\balance

\end{document}